# Spatially Resolved Mapping of Local Polarization Dynamics in an Ergodic Phase of Ferroelectric Relaxor

S.V. Kalinin,[*] B.J. Rodriguez, J.D. Budai, and S. Jesse

Oak Ridge National Laboratory, Oak Ridge, TN 37831

A.N. Morozovska

V. Lashkaryov Institute of Semiconductor Physics, National Academy of Science of Ukraine,

41, Prospect, Nauki, 03028 Kiev, Ukraine

A.A. Bokov and Z.-G. Ye

Department of Chemistry and 4D LABS, Simon Fraser University,

Burnaby, British Columbia, Canada

Spatial variability of polarization relaxation kinetics in relaxor ferroelectric $0.9Pb(Mg_{1/3}Nb_{2/3})O_3$-$0.1PbTiO_3$ is studied using time-resolved Piezoresponse Force Microscopy. Local relaxation attributed to the reorientation of polar nanoregions is shown to follow stretched exponential dependence, $\exp(-(t/\tau)^\beta)$, with $\beta \approx 0.4$, much larger than the macroscopic value determined from dielectric spectra ($\beta \approx 0.09$). The spatial inhomogeneity of relaxation time distributions with the presence of 100-200 nm "fast" and "slow" regions is observed. The results are analyzed to map the Vogel-Fulcher temperatures on the nanoscale.

---

[*] sergei2@ornl.gov

The unique electromechanical and dielectric properties of relaxor ferroelectrics have made them the materials of choice for numerous industrial and medical applications.[1] At the same time, a gamut of complex temperature-dependent dynamic behaviors and phase transitions in relaxors constitute one of the most challenging subjects in the physics of ferroelectrics.[2] At high temperatures, both relaxors and ferroelectrics exist in a non-polar paraelectric state. Below the Curie temperature, ferroelectrics transform into a ferroelectric phase, while relaxors undergo a transition to an ergodic relaxor state at the Burns temperature near which dynamic polar nanoregions (PNRs) with random dipole moment directions appear.[3] With decreasing temperature, the dynamics of PNRs slows down until they become frozen and the relaxor transforms into a nonergodic state that lacks long-range ferroelectric order and resembles a dipolar glass state. Alternatively, the transition from ergodic relaxor to ferroelectric state may occur in some crystals.[4]

The intrinsic link between the PNRs and unusual dynamic properties of relaxors has stimulated a number of spatially-resolved studies of mesoscopic polarization distributions using Piezoresponse Force Microscopy (PFM).[5,6] In PFM, the detection of local surface displacements induced by a periodically biased probe tip allows mapping of local electromechanical response and thereby reveals the domain structure, typically with ~10-30 nm spatial resolution. A number of studies[7,8] have reported the presence of nanoscale domains in relaxor materials such as $(1-x)Pb(Mg_{1/3}Nb_{2/3})O_3$ -$x$ $PbTiO_3$ and $(1-x)Pb(Zn_{1/3}Nb_{2/3})O_3$ -$x$ $PbTiO_3$. These domain structures were observed to be stable both below and well above (~100K) the Curie temperature $T_c$.[9] Even though the spatial resolution is significantly larger than the estimated size of the PNR (2-10 nm), these *static* studies provided insights into the relationship between disorder and mesoscopic (~100 nm) domain structures.

Despite this progress, a number of outstanding questions remain unanswered. Among them is the unique *dynamic* behavior in ferroelectric relaxors. This behavior has been investigated using macroscopic techniques such as dielectric spectroscopy[10,11] and light scattering[12,13] as well as by NMR spin lattice relaxation.[14] Macroscopically, relaxors demonstrate stretched exponential or fractional power law dynamics, consistent with a broad distribution of relaxation times.[11,15] However, the variation of local degrees of freedom of relaxation in space has not been studied. Thus the natural question is whether the relaxation time distribution associated with distinct physical elements (e.g. flipping PNRs) follows the Debye relaxation law, or the dynamics of each PNR is non-Debye due to interactions with other PNRs and built-in random fields. This question is relevant not only in the field of relaxors, but also in disordered materials in general.

This Letter reports on the direct measurements of the spatial distribution of relaxation parameters on the nanoscale in disordered materials. We study relaxor $0.9Pb(Mg_{1/3}Nb_{2/3})O_3$-$0.1PbTiO_3$ (PMN-10PT) crystals using spatially and time-resolved Piezoresponse Force Microscopy. We find nanoscale heterogeneity of the relaxation process, quantitatively characterize this heterogeneity and discuss its possible mechanisms.

The PMN-10PT crystal studied is grown from high temperature solution as described elsewhere.[16] The dielectric maximum occurs at $T_{max}$ = 310 K (at 1 kHz). The crystal undergoes a cubic relaxor to rhombohedral ferroelectric phase transition during cooling at $T_c \cong 280$ K.[17] The Burns temperature is ~ 650 K.[21] The absence of macroscopic piezoelectric effects[18] and aging[19] suggests that the room-temperature state in PMN-10PT is ergodic. The surface crystallography was verified using a combination of focused monochromatic and polychromatic depth-resolved x-ray experiments on beamline 34-ID at the Advanced Photon

Source as described by Larson et al.[20] Shown in Fig 1(a) is the (006) beam diffracted intensity as a function of depth using a monochromatic microbeam (E=12.307 keV). The peak intensity lies along a horizontal line, indicating a constant value for the *c*-lattice parameter within the error bar of $\Delta d/d=\pm 10^{-4}$. This data is also plotted in Fig 1(b), along with results from fitting depth-resolved, polychromatic (8-23 keV) Laue diffraction patterns (strain resolution $\Delta d/d \sim 10^{-4}$ and angle $\Delta\alpha \sim 0.03°$). These data demonstrate that the average crystal lattice remains cubic from the surface down to ~20 microns below the surface, consistent with the assumption that the measured region is in the ergodic relaxor phase.

The PFM measurements are performed using a commercial AFM (Veeco MultiMode with Nanonis controller) on the mirror-polished (001) cut of the crystal. Typical topography and domain patterns are shown in Fig. 1 (c-e). The presence of switchable polarization was established using a PFM switching experiment [Fig. 1 (f-h)]. Macroscopic relaxation behavior is studied using dielectric spectroscopy[11] and the results are summarized in Fig. 2 (a). To probe *local* relaxation behavior, DC bias pulses of specified magnitude and duration are applied to the conducting AFM tip in contact with the sample, and the resulting vertical electromechanical response is measured as a function of time for a specified duration. To prove that the detection signal does not affect the relaxation, measurements are performed for several $V_{ac}$ amplitudes (1-3, 10 $V_{pp}$) and the relaxation is found to be essentially similar.

The time-resolved spectroscopic measurements are performed in the *dense* regime where the pixel spacing is significantly smaller than the spatial extent of material affected by the field. The relaxation curves at each pixel are fitted to a specific relaxation law, $R = R_0 + R_1 f(t)$. Here, $R$ is the measured PFM signal. The offset, $R_0$, and amplitude, $R_1$, correspond to non-relaxing (within the measurement time scale) and relaxing polarization components, respectively. The function $f(t)$ is chosen as Debye, $f(t) = \exp(-t/\tau_D)$, stretched exponential or Kolrausch-William-Watts (KWW), $f(t) = \exp(-(t/\tau_{KWW})^\beta)$, fractional power or Curie-von Schweidler (CvS), $f(t) = (t/\tau_{CvS})^n$, or logarithmic, $f(t) = \ln(t)$. The relationship between the relaxation laws and activation energy distributions has been extensively studied in the context of macroscopic theories.[21] After least-square fitting, the model-dependent parameters (e.g. $R_0$, $R_1$, $\tau_{KWW}$, and β for stretched exponential fit) are plotted as 2D maps that can be further correlated with local microstructure.

To determine the long-term relaxation dynamics, single-point relaxation data are acquired in the time interval from 10 ms to 100s. Fig. 2 (b) demonstrates the typical decay of piezoresponse in one of the crystal surface points measured after dc bias pulse. Over 4 orders of magnitude, the relaxation can be well-described by the KWW law with $\beta = 0.49$, $\tau_{KWW} = 15$ s. The simple exponential, logarithmic and CvS fits all fail. Notably, macroscopic (dielectric) measurements reveal the KWW behavior with $\beta = 0.09$.

To quantitatively analyze these observations, we consider standard Debye-like relaxation dynamics $dP/dt = -P/\tau$, where $\tau$ is the relaxation time. Assuming a distribution of relaxation times described by normalized distribution function $g(\tau)$, the average response is

$$\langle P \rangle = P_0 \int_0^\infty d\tau \cdot g(\tau) \exp\left(-\frac{t}{\tau}\right) \tag{1}$$

The distribution functions $g(\tau)$ could be reconstructed from experimentally measured values using inverse Laplace transformation. The relaxation time distributions corresponding to the KWW data from Fig. 2(a,b) are shown in Figs. 2 (c,d). For this functional form, the numerical analysis reveals two well-defined regimes. For $\beta > 0.5$ the sharp peak appears in $g(\tau)$, corresponding to the system with a "smeared" Debye relaxation. For $\beta < 0.5$, the spectral weight shifts to small relaxation times and $g(\tau) \approx \tau^{-1}$. The fine details of the distribution function are better seen from $G(\ln \tau) \equiv \tau g(\tau)$ [Fig.2 (c-d)]. The limiting case of logarithmic relaxation corresponds to a KWW distribution function with $G(\ln \tau) \approx \text{const}$.

To study spatial variability of the relaxation behavior the measurements are performed on a closely (50 nm) spaced 40 x 40 grid. A setting pulse of 10 V amplitude is applied to the probe for 30 ms, and then the bias is turned off for the following 300 ms. The results are averaged over 3 repetitions. The electromechanical response (measured at 1.1 MHz) is fitted using the KWW model. Thus derived parameter maps are shown in Fig. 3. In the writing process [Figs. 3(a-c)] the switchable polarization, $R_1$, shows large-scale features associated with strong contrast variation within the image (~30%), partially associated with topographic details [Fig. 1 (f)]. At the same time, the spatial maps of relaxation time, $\tau_{KWW}^{W}$, and exponent, $\beta$, are generally featureless, with effective noise level higher than the large-scale contrast.

The 2D maps corresponding to zero-field relaxation (reading) illustrate different dynamics [Figs. 3(d-f)]. The relaxation amplitude image shows the pronounced contrast similar to that observed on writing. The relaxation time, $\tau$, image in Fig. 3(e) illustrates the presence of "slow" and "fast" regions on the length scale of 100-200 nm. The response time differs by a factor of 4. Similarly, $\beta$ images show large spatial variability, with exponent changing between 0.5 and 0.3, depending on position. Both visual inspection and cross-correlation analysis suggest that the 2D maps contain complementary information on local properties. This indicates that the amount of polarization that relaxes within the time interval of measurements is clearly position dependent.

The statistical distributions for some of the parameters are shown in Figs. 3 (g,h). The relaxation time distributions for both writing and reading processes are rather broad. An order of magnitude difference in writing and reading relaxation times is expected and related to the fact that comparatively large dc bias field is applied during the writing process, which effectively reduces the activation energy for the relaxation. The values of $\tau_{KWW}$ in reading measurements are significantly smaller than in the single-point experiment in Fig. 2 (b). This is because the duration of dc bias pulse in the former case (30 ms) is not long enough to excite the long-time degrees of freedom and consequently the relaxation spectrum is cut from the long-time side. The values of $\beta$ of ~ 0.4 for the reading process correspond to a virtually flat distribution of relaxation times. It should be emphasized the dimensions of static labyrinthine domains and dynamic fast and slow regions as well as the size of regions probed in single-point experiment (~30 nm) are much larger than the expected PNR size (~2 nm).[22] Hence, single-point data are the result of averaging over several PNRs.

To interpret the experimental results we use a model assuming the existence of two types of PNRs in an ergodic phase of relaxors: static and dynamic.[13,23] The static PNRs are responsible for the formation of labyrinthine domains (frozen polarization fluctuations) existing before the application of the external field [Fig. 1 (d)] while dynamic PNRs give rise to the observed relaxation. The writing dc pulse triggers the reorientations of dynamic PNRs and consequently the appearance of additional long-lived polarization, $R_1$. After switching off the dc field, dynamic PNRs relax to the state with a random distribution of dipole moments causing the KWW-type decrease of PFM signal. The spatial variation of relaxation parameters is related partially to the topographic features of the surface and partially to the random interactions among dynamic PNRs and random fields caused by quenched disorder and by static PNRs.[23,24,25,26] Note that recent NMR experiments also confirmed the existence in nonergodic relaxor phase of static (on the $10^{-4}$ s scale) PNRs along with the dynamic ones. The PFM reading relaxation rate corresponds to the rate of dielectric relaxation related to the reorientation of dynamic PNRs. The macroscopic dielectric relaxation time of ~0.2 s [Fig. 2 (a)] is indeed close to the mean PFM reading $\tau_{KWW}$ (~0.1 s). In contrast, the PFM values of $\beta$ ~0.4 are the local parameters which characterize relaxation time distribution inside the probed nanoscale regions [Fig. 3 (h)] with much fewer degrees of freedom. Therefore, the macroscopic value of $\beta$ is expected to be smaller, as observed (the dielectric $\beta$ is ~0.09).

To relate the materials parameters to the relaxation law we assume that the local relaxation time depends on activation energy, $E$, in accordance with the Vogel-Fulcher relationship, $\tau(E) = \tau_0 \exp(E/(T - T_f))$. Then the relaxation law in terms of distribution function of energies $G(E)$ is expressed as

$$\langle P \rangle = P_0 \int_{E_{min}}^{E_{max}} dE\, G(E) \exp\left(-\frac{t}{\tau(E)}\right). \tag{2}$$

Eq.(2) leads to $G(E) \equiv (T - T_f)^{-1} \tau(E) g(\tau(E))$. When the energy distribution is almost uniform, $G(E) \approx (E_{max} - E_{min})^{-1}$, Eq. (2) can be integrated in the analytical form.[27] For $\tau_{min} \ll t \ll \tau_{max}$, where $\tau_{min,max} = \tau_0 \exp(E_{min,max}/(T - T_f))$, we obtain

$$\langle P(t) \rangle \approx -P_0 \left( \frac{T - T_f}{E_{max}} \gamma - 1 + \frac{T - T_f}{E_{max}} \ln\left(\frac{t}{\tau_0}\right) \right) \tag{3}$$

where the Euler constant $\gamma = 0.577$. Hence, the spatial variation of slope in response-time dependence for the logarithmic model can be interpreted as the fluctuations of the local Vogel-Fulcher temperature, as shown in Fig. 4 (a-c).

To summarize, we have studied the ergodic relaxor phase of PMN-10PT crystal in which the relaxation is attributed to the reorientation of dynamic PNRs. The measurements over 4 orders of magnitudes in time indicate that local relaxation dynamics follows a stretched exponential function, with the observed values of $\beta$ corresponding to an almost uniform distribution of activation barrier heights. The relaxation spectrum in the macroscopic sample are much wider than the local spectrum due to averaging over mesoscopic spatial inhomogeneities. The local relaxation kinetics is more Debye-like, corresponding to the small number of PNRs within the probing volume of a PFM tip. Significant spatial distribution of relaxation behaviors linked to the internal fields and surface topographic variations is observed. We note that local non-Debye relaxation dynamics is common to many systems with structural (glasses, polymers), magnetic (spin glasses) or polar (dipole glass) disorder. The strong coupling between polarization and strain (reversible lattice deformation) in relaxors and reversibility of switching behavior allows us to study the dynamics locally using PFM and makes relaxors an ideal model for studying general relaxation principles in disordered systems including structural glasses, polymers (in which dynamics is irreversible) and spin- and cluster glasses (in which mapping local magnetization is a challenge).

This research is supported by the Division of Scientific User Facilities, U.S. DOE. X-ray work at XORUNI-APS sponsored by DMSE USDOE BES under contract with UT-Battelle. The work is also supported (AAB, ZGY) by the U. S. Office of Naval Research (Grant No.N00014-06-1-0166). The authors thank W. Chen and X. Long for help in crystal preparation and J. Tischler and W. Liu for help with x-ray measurements. BJR gratefully acknowledges the Alexander von Humboldt foundation for financial support.

**Figure Captions**

**Fig. 1.** (a) Monochromatic x-ray diffraction intensity as a function of scattering vector, **Q**, and depth below the sample surface. (b) Lattice parameters extracted from the monochromatic (squares) and polychromatic Laue (triangles) x-ray microdiffraction measurements as a function of depth. Vertical lines correspond to error bars in mono (blue) and polychromatic measurements. (c,f) Surface topography; (d,g) PFM amplitude; and (e,h) PFM phase images of PMN-10PT surface. (c,d,e) Pristine domains structure. (f,g,h) Changes in domain structure after time-resolved PFM mapping (different location).

**Fig. 2.** (a) Frequency dependencies of real ($\varepsilon'$) and imaginary ($\varepsilon''$) parts of permittivity measured in (001) oriented PMN-10PT crystal at 21 $^{\circ}$C. Solid lines represent fitting to frequency-domain transform of KWW function with the parameters $\tau_{KWW}$ = 0.2 s and $\beta$ =0.09. (b) Vertical piezoresponse measured after switching off the dc bias signal of 10 V applied for 100 ms and fitted by different relaxation laws: exponential, CvS and KWW. Best fit corresponds to the KWW relaxation law with $R_0$ =0.65; $R_1$=2.85 $\beta$ = 0.49; $\tau_{KWW}$=15 s. (c, d, e) KWW relaxation time $\tau$ and activation energy $E$ distribution functions $g(\tau)$, $G(\ln \tau)$ and $G(E)$ for different β (labels near the curves, β=0.49 corresponds to experiment data).

**Fig. 3.** Spatially-resolved mapping of polarization dynamics in PMN-10PT using stretched exponential law. KWW parameters for (a,b,c) writing and (d,e,f) reading processes. (g,h) Histograms of relaxation time and exponents.

**Fig. 4.** (a) Nearly logarithmic relaxation in small time interval 0.1-10s measured after dc bias signal of different durations. (b,c) Spatially-resolved mapping of polarization dynamics in PMN-10PT using logarithmic decay. Shown are (b) intercept and (c) slope for reading.

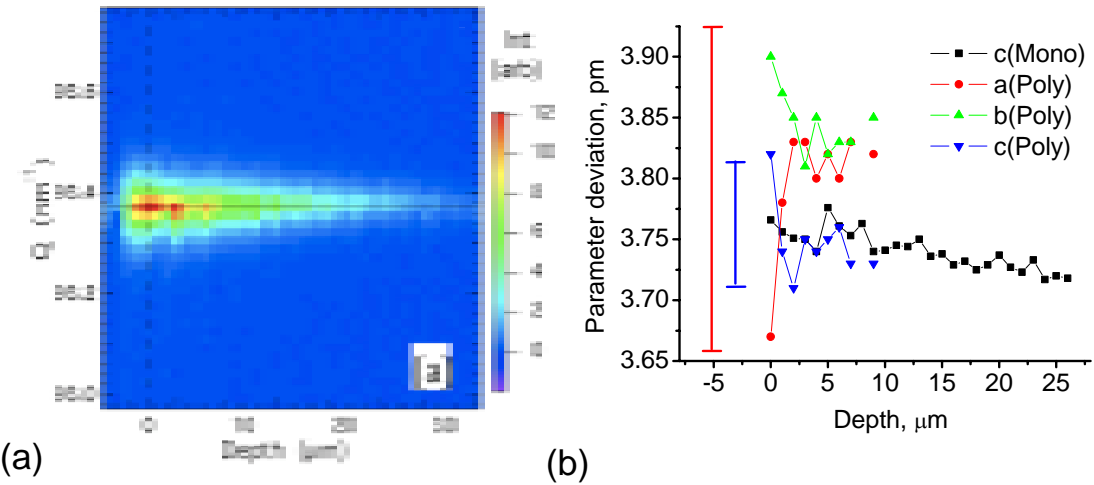
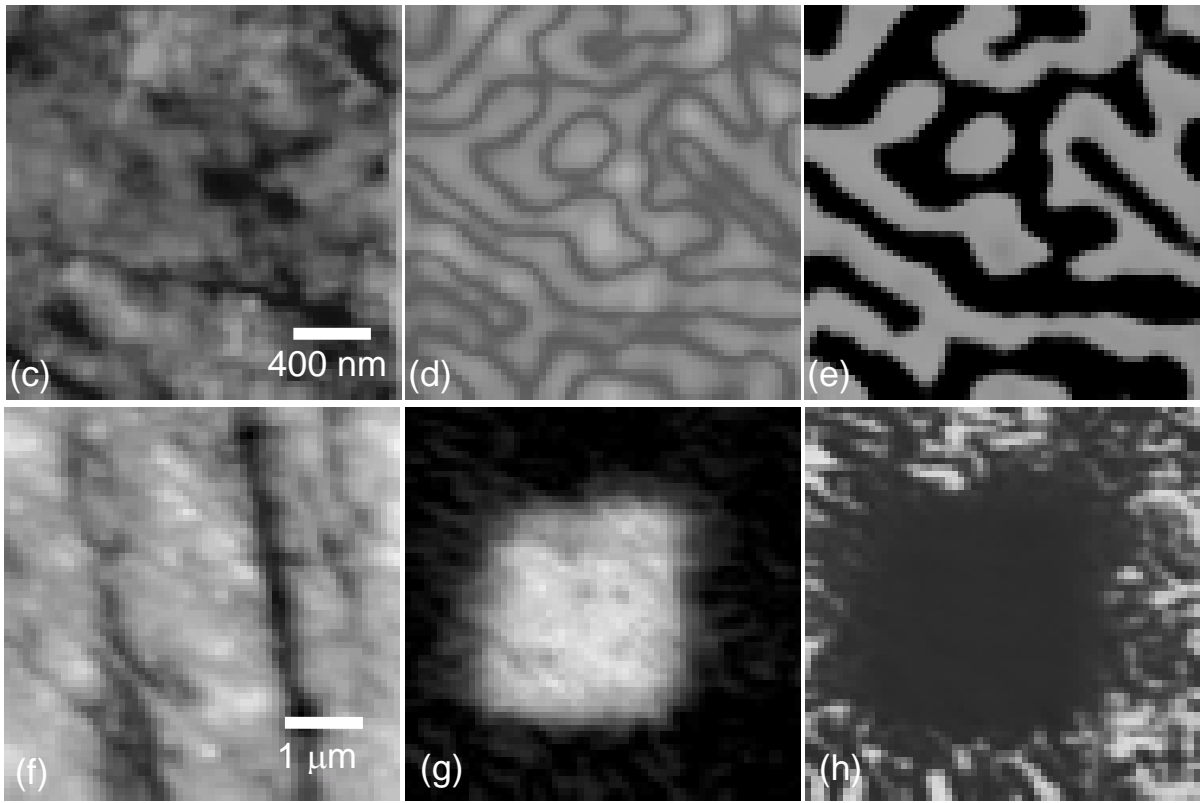

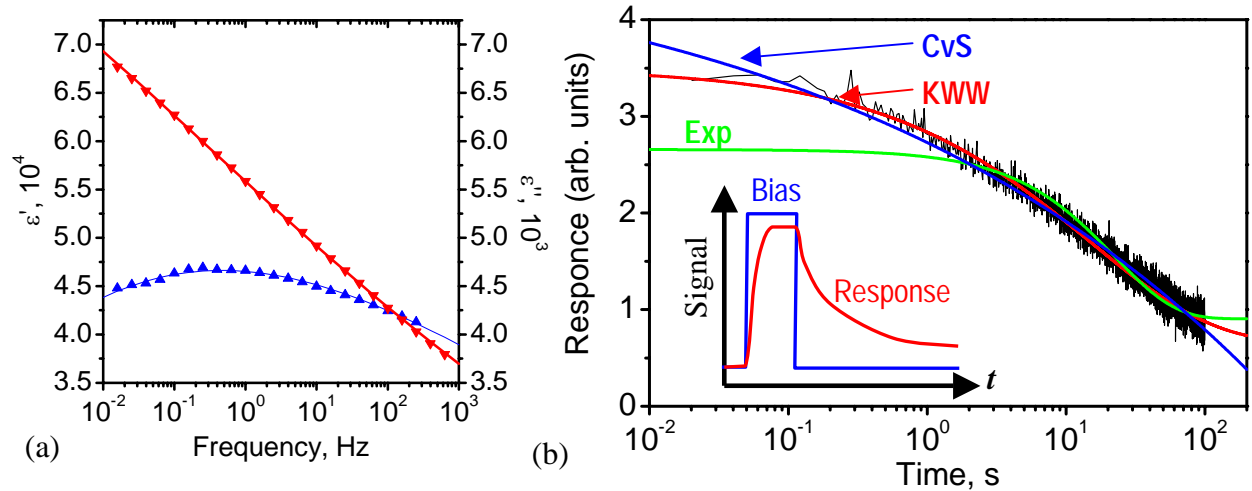
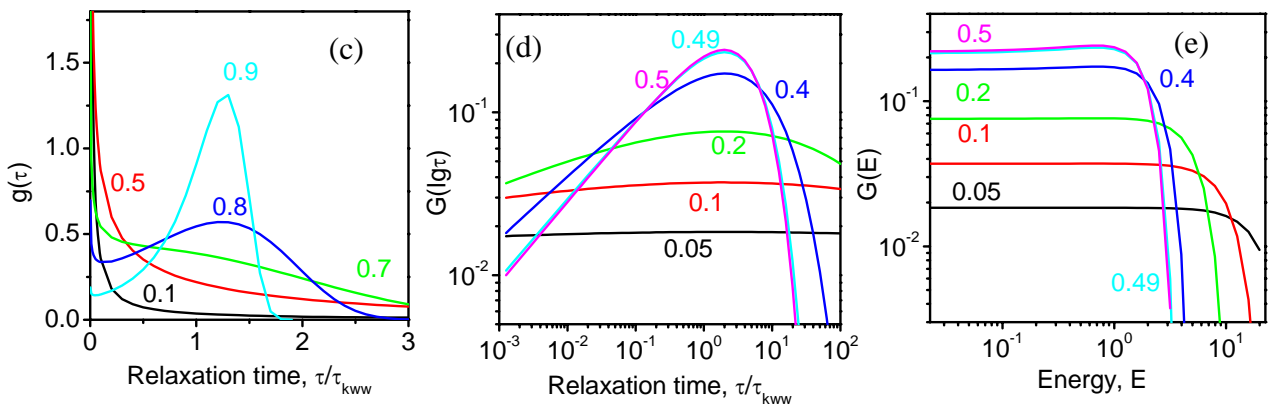

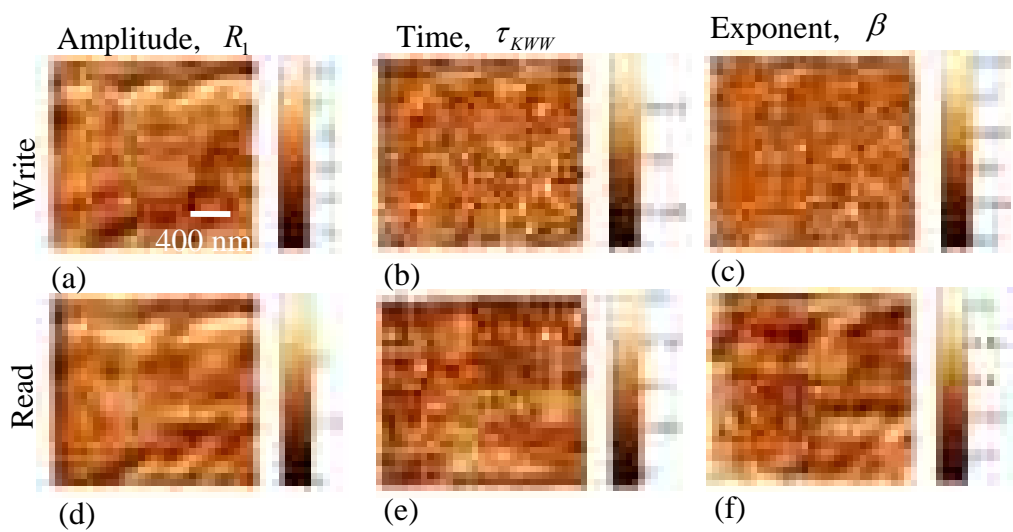

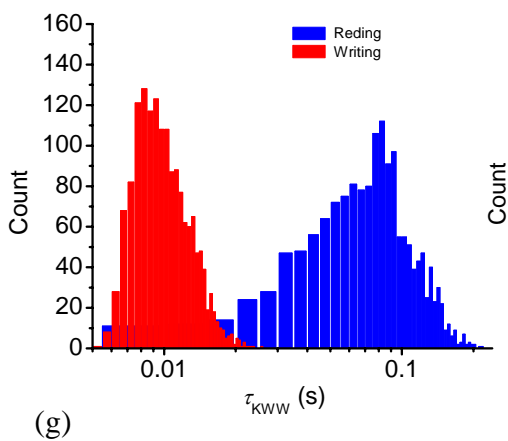 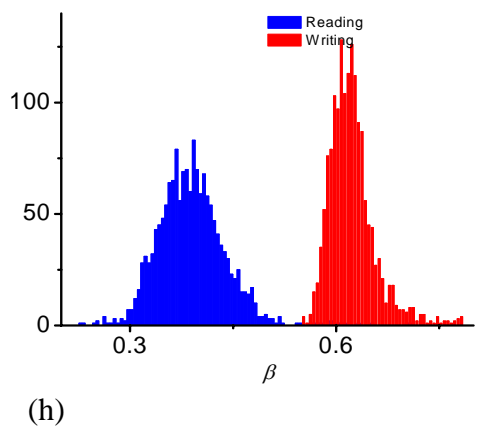

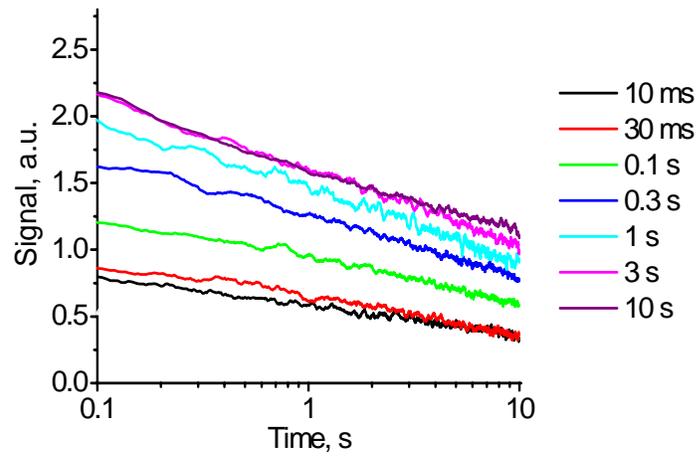

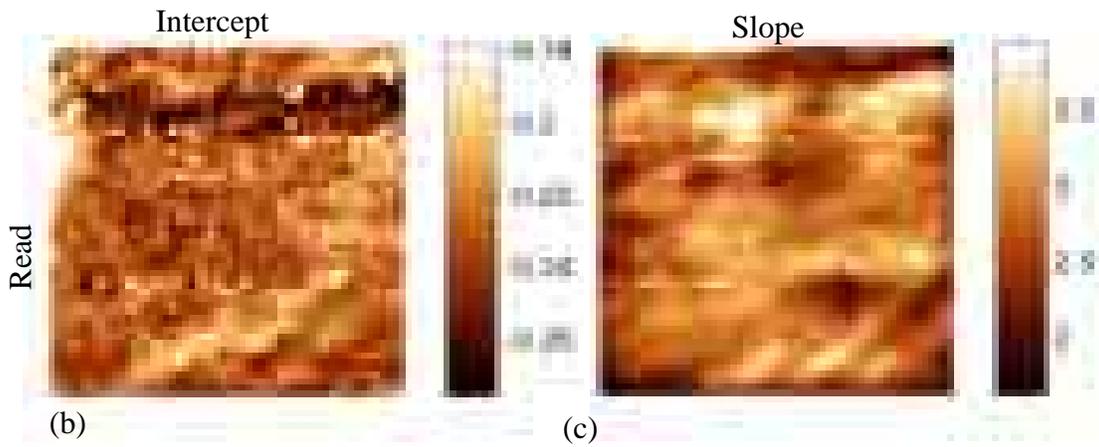

(a)

(b) Intercept

(c) Slope

Read

# References

[1] W. Kleemann, J. Mat. Sci. **41**, 129 (2006).

[2] A.A. Bokov, Z.-G. Ye, J. Mat. Sci. **41**, 31 (2006).

[3] G. Burns and F.H. Dacol, Phys. Rev. **B 28**, 2527 (1983).

[4] B. Dkhil, J.M. Kiat, G. Calvarin, G. Baldinozzi, S. B. Vakhrushev, and E. Suard, Phys. Rev. **B 65**, 024104 (2001).

[5] M. Abplanalp et al, Solid State Comm. **119,** 7 (2001).

[6] A. Gruverman and A. Kholkin, Rep. Prog. Phys. **69**, 2443 (2006).

[7] V.V. Shvartsman, et al, Phys. Rev. **B 69**, 014102 (2004).

[8] V.V. Shvartsman, et al, Appl. Phys. Lett. **86**, 202907 (2005).

[9] V.V. Shvartsman and A.L. Kholkin, J. Appl. Phys. **101**, 064108 (2007).

[10] A. Levstik, Z. Kutnjak, C. Filipic, and R. Pirc, Phys. Rev. **B 57,** 11204 (1998).

[11] A.A. Bokov and Z.G. Ye, Phys. Rev. B **74**, 132102 (2006).

[12] J.-H. Ko, D. H. Kim, and S. Kojima, Phys. Rev. B **77**, 104110 (2008).

[13] J.-H. Ko, S. Kojima, A.A. Bokov, and Z.-G. Ye, Appl. Phys. Lett. **91**, 252909 (2007).

[14] R. Blinc, V. Laguta, and B. Zalar, Phys. Rev. Lett. **91**, 247601 (2003).

[15] W. Kleemann and R. Linder, Ferroelectrics **199**, 1 (1997).

[16] M. Dong and Z.-G. Ye, J. Crystal Growth **209**, 81 (2000).

[17] Z.-G. Ye, Y. Bing, J. Gao, A.A. Bokov, P. Stephens, B. Noheda, and G. Shirane, Phys. Rev. **B 67**, 104104 (2003)

[18] W.Y. Pan, W.Y. Gu, D.J. Taylor, and L.E. Cross, Jpn. J. Appl. Phys. **28,** 653 (1989).

[19] D. Viehland, J.F. Li, S.J. Jang, L.E. Cross, and M. Wittig, Phys. Rev. B **43**, 8316 (1991);

[20] B. Larson et al, Nature **415**, 887 (2002)

[21] R. Hilfer, Journal of Non-Crystalline Solids **305,** 122 (2002).

[22] G. Xu, G. Shirane, J.R.D. Copley, and P.M. Gehring. Phys. Rev. B **69,** 064112 (2004).

[23] X. Long, A.A. Bokov, Z.-G. Ye, W. Qu, and X. Tan, J. Phys.: C **20**, 015210 (2008).

[24] V. Westphal, W. Kleemann, and M.D. Glinchuk, Phys. Rev. Lett. **68**, 847 (1992).

[25] R. Pirc and R. Blinc, Phys. Rev. B **60,** 13470 (1999).

[26] R. Fisch, Phys. Rev. B **67,** 094110 (2003).

[27] Supplementary Materials

# Spatially Resolved Mapping of Local Polarization Dynamics in an Ergodic Phase of Ferroelectric Relaxor

S.V. Kalinin,[1] B.J. Rodriguez, J.B. Budai, and S. Jesse

Oak Ridge National Laboratory, Oak Ridge, TN 37831

A.N. Morozovska

V. Lashkaryov Institute of Semiconductor Physics, National Academy of Science of Ukraine, 41, Prospect, Nauki, 03028 Kiev, Ukraine

A.A. Bokov and Z.-G. Ye

Simon Fraser University, Department of Chemistry, Burnaby, British Columbia, Canada

## I. Surface state analysis

Similar to pure PMN and $Pb(Zn_{1/3}Nb_{2/3})O_3$ relaxors, the low-temperature rhombohedral phase is observed (by means of x-ray diffraction) only in the outer ~50 micron layer of the crystal; neutron diffraction experiments (which sample the crystal bulk) reveal an internal ~cubic phase.[1] Thus, at room temperature where the experiments are carried out, the PMN-10PT crystal (in both the bulk and on the surface) is in the cubic ergodic relaxor phase. Note that due to the closeness of diffuse ferroelectric transition nonergodic effects may occur, although they have not been reported. On the other hand, the absence of macroscopic piezoelectric effects[2] and aging[3] suggests that the room-temperature state in PMN-10PT is still ergodic.

---

[1] sergei2@ornl.gov

In our study, the surface crystallography was verified using a combination of focused monochromatic and polychromatic depth-resolved x-ray experiments on beamline 34-ID at the Advanced Photon Source as described recently by Larson et al.[4] Shown below are unit cell parameters from monochromatic and polychromatic data (similar to Fig. 1(b) in the main manuscript) and the corresponding angles.

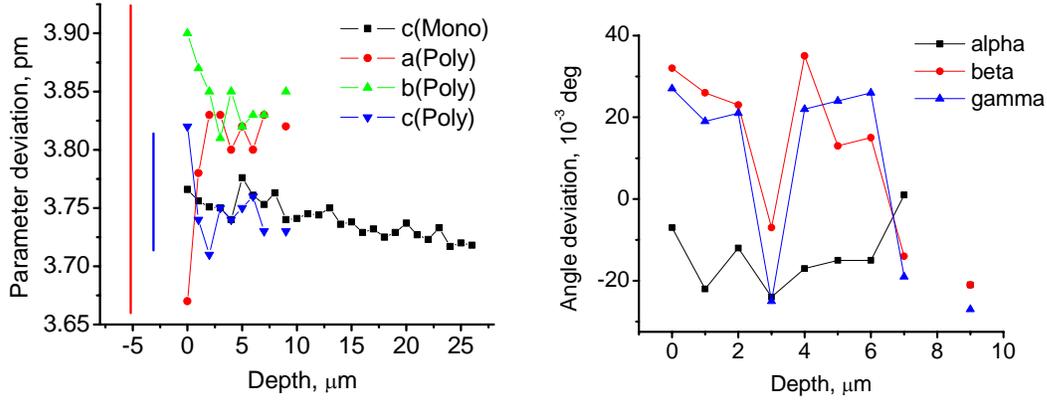

Fig. S1. Depth-resolved focused X-ray study of PMN-10PT surface.

## II. Relaxation Dynamics

We consider standard relaxation dynamics $\frac{dP}{dt} = -\frac{P}{\tau}$, which is equivalent to relaxation. $P = P_0 \exp\left(-\frac{t}{\tau}\right)$. We assume that relaxation time depends on local potential energy $E$ in accordance with Vogel-Fulcher relationship, namely $\tau(E) = \tau_0 \exp\left(\frac{E}{T - T_f}\right)$.

Assuming the distribution of relaxation times described by normalized distribution function $g(\tau)$, the average response is

$$\langle P \rangle = P_0 \int_0^\infty d\tau \cdot g(\tau) \exp\left(-\frac{t}{\tau}\right) \tag{1}$$

In accordance with Vogel-Fulcher relationship $\frac{d\tau}{dE} = \frac{\tau(E)}{T-T_f}$. Then the relaxation law in distribution function of energies $g(E)$ is

$$\langle P \rangle = P_0 \int_0^\infty d\tau \cdot g(\tau) \exp\left(-\frac{t}{\tau}\right) = \left| d\tau = \frac{\tau(E)dE}{T-T_f} \right| = P_0 \int_{E_{min}}^{E_{max}} dE \frac{\tau(E)}{T-T_f} g(\tau(E)) \exp\left(-\frac{t}{\tau(E)}\right) \tag{2}$$

The relaxation time distribution for KWW law is $g(\tau) = \frac{1}{\tau} \sum_{n=0}^\infty \frac{(-1)^n}{n!\Gamma(-\beta n)} \left(\frac{\tau}{\tau_{KWW}}\right)^{\beta n}$

Eq.(2) immediately leads to $G(E) \equiv \frac{\tau(E)}{T-T_f} g(\tau(E))$ by definition, so that when $G(E) \approx (E_{max} - E_{min})^{-1}$ is constant one obtains.

$$\langle P \rangle \approx \frac{P_0}{E_{max} - E_{min}} \int_{E_{min}}^{E_{max}} dE \exp\left(-\frac{t}{\tau_0} \exp\left(-\frac{E}{T-T_f}\right)\right) \tag{3}$$

Exact integration again leads to

$$\langle P \rangle = \frac{P_0(T-T_f)}{E_{max} - E_{min}} \left( Ei\left(-\frac{t}{\tau_0} \exp\left(-\frac{E_{min}}{T-T_f}\right)\right) - Ei\left(-\frac{t}{\tau_0} \exp\left(-\frac{E_{max}}{T-T_f}\right)\right) \right) \tag{4}$$

where $Ei(z) = -\int_{-z}^{\infty} dy \exp(-y)/y$ is well-known (tabulated) exponential integral function, at that $-Ei(-z \to -\infty) \to \exp(-z)/z$ and $Ei(-z \to 0) \to \gamma + \ln z - z$, where Euler constant $\gamma = 0.577$.[5] Series expansion of Eq.(4) leads to the following limiting cases:

At $t \ll \tau_{min}$ ($\tau_{min,max} = \tau_0 \exp\left(\dfrac{E_{min,max}}{T - T_f}\right)$) we obtain

$$\langle P \rangle \approx \dfrac{P_0(T - T_f)}{E_{max} - E_{min}} \left( \begin{array}{l} \ln\left(\dfrac{t}{\tau_0}\exp\left(-\dfrac{E_{min}}{T - T_f}\right)\right) - \ln\left(\dfrac{t}{\tau_0}\exp\left(-\dfrac{E_{max}}{T - T_f}\right)\right) + \\ -\dfrac{t}{\tau_0}\exp\left(-\dfrac{E_{min}}{T - T_f}\right) + \dfrac{t}{\tau_0}\exp\left(-\dfrac{E_{max}}{T - T_f}\right) \end{array} \right) \qquad (5a)$$

or $\langle P \rangle \approx P_0 + \dfrac{t}{\tau_0}\left(\exp\left(-\dfrac{E_{max}}{T - T_f}\right) - \exp\left(-\dfrac{E_{min}}{T - T_f}\right)\right)$

At intermediate times $\tau_{min} \ll t \ll \tau_{max}$ ($\tau_{min,max} = \tau_0 \exp\left(\dfrac{E_{min,max}}{T - T_f}\right)$) we obtain

$$\langle P \rangle \approx -\dfrac{P_0(T - T_f)}{E_{max}}\left(\gamma + \ln\left(\dfrac{t}{\tau_0}\exp\left(-\dfrac{E_{max}}{T - T_f}\right)\right)\right) \text{ or }$$

$$\langle P \rangle \approx -P_0\left(\dfrac{T - T_f}{E_{max}}\gamma - 1 + \dfrac{T - T_f}{E_{max}}\ln\left(\dfrac{t}{\tau_0}\right)\right) \qquad (5b)$$

Finally, at $t \gg \tau_{max}$

$$\langle P \rangle \approx \dfrac{P_0(T - T_f)}{E_{max}}\exp\left(-\dfrac{t}{\tau_0}\exp\left(-\dfrac{E_{max}}{T - T_f}\right) + \dfrac{E_{max}}{T - T_f}\right)\dfrac{\tau_0}{t} \qquad (5c)$$

# References

[1] P. M. Gehring, W. Chen, Z.-G. Ye, and G. Shirane, J. Phys.: Condens. Matter **16**, 7113 (2004)

[2] W.Y. Pan, W.Y. Gu, D.J. Taylor, and L.E. Cross, Jpn. J. Appl. Phys. **28,** 653 (1989).

[3] D. Viehland, J.F. Li, S.J. Jang, L.E. Cross, and M. Wittig, Phys. Rev. B **43**, 8316 (1991);

[4] B. Larson et al, Nature **415**, 887 (2002)

[5] I.S. Gradshteyn and I.M. Ryzhik, Table of Integrals, Series, and Products, 5th ed., edited by A.Jeffrey (Academic, New York, 1994).